\documentclass[fleqn,usenatbib]{mnras}

\usepackage[T1]{fontenc}
\usepackage{ae,aecompl}

\usepackage{graphicx}	
\usepackage{amsmath}	
\usepackage{amssymb}	
\usepackage{multicol}
\usepackage{bm}
\usepackage{booktabs}
\usepackage{xcolor}
\usepackage{soul}
\usepackage{hyperref}

\usepackage{natbib}
\title[Non-Negative Matrix Factorization for Improving OSIRIS Calibration]{Using Non-Negative Matrix Factorization to Improve Calibration of the Keck OSIRIS Integral Field Spectrograph}

\author[K. Horstman et al.]{Katelyn Horstman$^{1,}$$^{2}$,\thanks{E-mail: khorstma@astro.caltech.edu}
Michael P. Fitzgerald$^{1}$,
James E. Lyke$^{3}$,\newauthor
Sherry C.~C. Yeh$^{3}$,
Devin S. Chu$^{1}$
\\
$^{1}$Department of Physics and Astronomy, University of California, Los Angeles, 430 Portola Plaza, Los Angeles, CA 90095, USA\\
$^{2}$Department of Astronomy, California Institute of Technology, Pasadena, CA 91125, USA\\
$^{3}$W. M. Keck Observatory, Waimea, HI 96743, USA\\
}

\begin{document}
\label{firstpage}
\maketitle

\begin{abstract}
Integral Field Spectrographs (IFS) often require non-trivial calibration techniques to process raw data. The OH Suppressing InfraRed Imaging Spectrograph (OSIRIS) at the W. M. Keck Observatory is a lenslet-based IFS that requires precise methods to associate the flux on the detector with both a wavelength and a position on the detector. During calibration scans, a single column lenslet mask is utilized to keep light from adjacent lenslet columns separate from the primary lenslet column, in order to uniquely determine spectral response of individual lenslets on the detector. Despite employing a single column lenslet mask, an issue associated with such calibration schemes may occur when light from adjacent masked lenslet columns leaks into the primary lenslet column. Incorrectly characterizing the flux due to additional light in the primary lenslet column results in one form of crosstalk between lenslet columns, which most clearly manifest as non-physical artifacts in the spectral dimension of the reduced data. We treat the problem of potentially blended calibration scans as a source separation problem and implement Non-negative Matrix Factorization (NMF) as a way to separate blended calibration scan spectra. After applying NMF to calibration scan data, extracted spectra from calibration scans show reduced crosstalk of up to 26.7$\pm$0.5$\%$ while not adversely impacting the signal-to-noise ratio. Additionally, we determined the optimal number of calibration scans per lenslet column needed to create NMF factors, finding that greatest reduction crosstalk occurs when NMF factors are created using one calibration scan per lenslet column.
\end{abstract}

\begin{keywords}
non-negative matrix factorization, crosstalk, spectral calibration, integral field spectrograph
\end{keywords}

\section{Introduction} 
\label{sec:intro}

Lenslet-based integral field spectrographs (IFSs) often require the use of non-trivial calibration methods to process raw data because of their unique design and components. For example, in order to maximize the effectiveness of the instrument, by expanding its field of view or wavelength coverage, one possible design choice is to pack as many spectra as possible on the detector.The first IFSs to use lenslet arrays, Traitement Intégral des Galaxies par l'Etude de leurs Raies (TIGER) on the Canada France Hawaii Telescope (CFHT) and the Multi-Pupil Integral Field Spectrograph (MPFS) on the Bolshoi Teleskop Alt-azimutalnyi (BTA-6), achieved first light the late 20th century \citep{bacon_1995, afanasiev_1995}. After the IFSs succeeded in obtaining both spatial and spectral information from astrophysical objects, the Spectroscopic Areal Unit for Research on Optical Nebulae (SAURON) on the William Herschel Telescope (WHT) was created to improve on previous designs by providing a wider field of view and higher spatial resolution. To gain a wider field of view, SAURON chose to maximize number of spectra on the detector by overlapping spectra by 10$\%$ \cite{bacon_2001}. 
Taken to an extreme, the partial overlap of spectra from individual lenslets can lead to the improper characterization of light on the detector.

Disentangling overlapping spectra is a common limitation associated with IFS calibration and is observed in seeing-limited IFSs such as TIGER, MPFS, and SAURON, and AO-fed IFSs such as the Gemini Planet Imager (GPI), the Coronagraphic High Angular Resolution Imaging Spectrograph (CHARIS), and the OH Suppressing InfraRed Imaging Spectrograph (OSIRIS) \citep{bacon_2001, in_2014, brandt_2017, lockhart2019}. Properly separating blended spectra to perform spectral extraction can be classified as a blind source separation problem. 

Spectral extraction requires deblending procedures that rely calibration techniques to assign the flux on the detector to a three-dimensional data cube consisting of both spatial and spectral dimensions. Calibration procedures use the line profiles of monochromatic emission sources, such as arc-lamp spectra, to determine a wavelength solution. To produce well-separated spectra for calibration procedures, individual lenslet columns in the lenslet array are isolated. The flux from each lenslet can then easily be assigned to individual pixels on the detector \citep{lockhart2019}.

\begin{figure}
\begin{centering}
\includegraphics[scale=0.3]{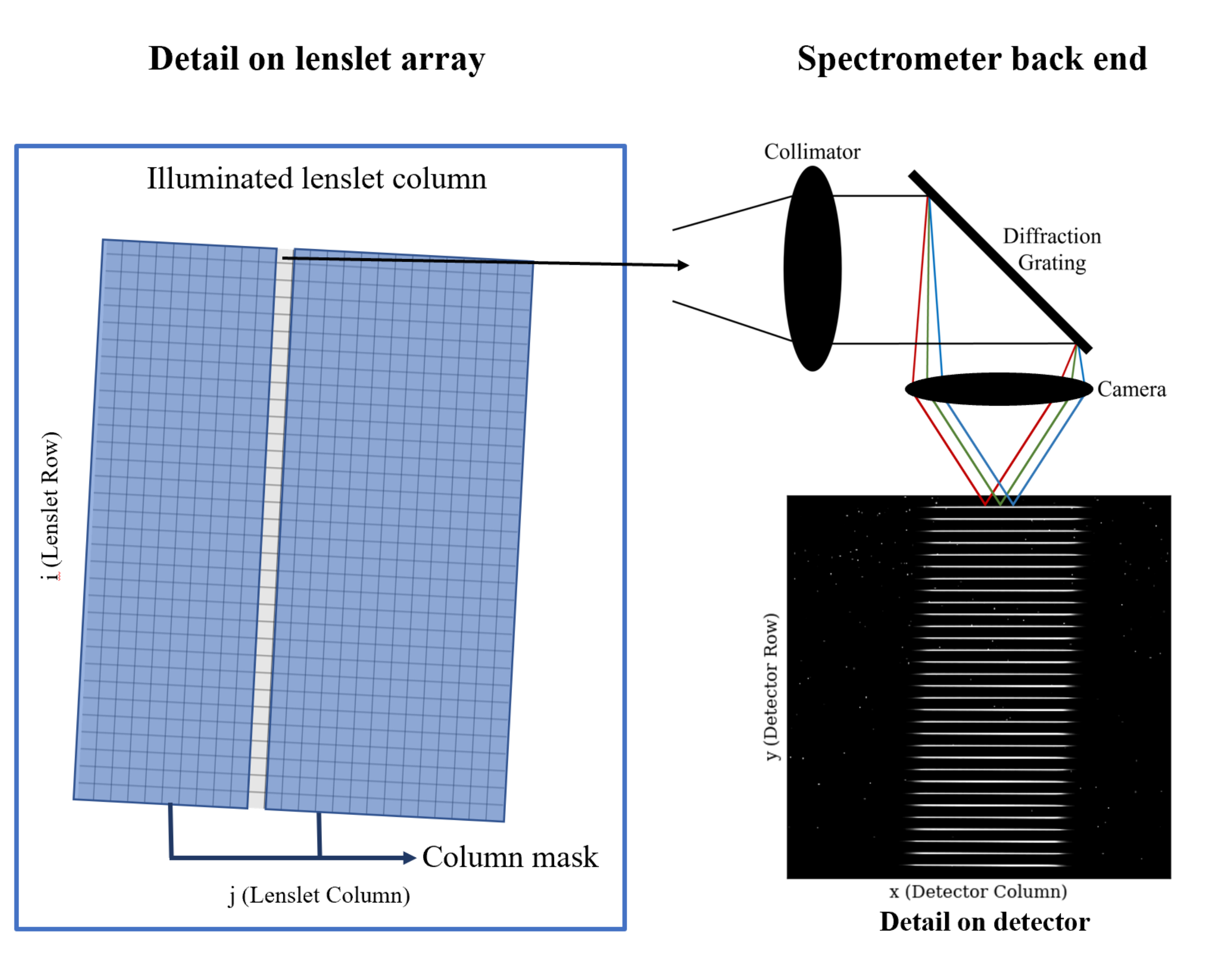}
\caption{Schematic of how light travels through a lenslet array to a detector. The schematic only depicts the upper left quarter of the illuminated lenslet column. The detail on the detector shows an example of the order trace.}
\label{fig:det_setup}
\end{centering}
\end{figure}

Using a single lenslet column mask on a motorized slide, calibration images can be taken at the position of a single lenslet column. Figure~\ref{fig:det_setup} shows the configuration of how light passes from the lenslet array, through the optical system, to the detector of a lenslet-based IFS during arc-lamp calibrations. If the mask is oversized, or imperfectly aligned with the lenslet columns, light from adjacent lenslet columns can leak into the primary lenslet column or light from primary lenslet column can scatter internal to the lenslet array and leak into adjacent lenslets. This causes calibration light to blend, impeding our ability to use calibration scans for deblending procedures. Mismatch between the ``true'' lenslet response on detector and the response caused by the blended calibration sources leads to flux mis-assignment. In processed images of arc-lamp spectra, blended calibration sources can manifest as dark, ``ghost'' spectral lines that are not really present in the image.
The negative flux associated with ``ghost'' spectral lines is one way to quantify crosstalk between adjacent lenslet columns. 

For calibration scans to be useful in deblending procedures, the calibration light itself cannot be blended, even though this is hard to guarantee in an experimental setup. One potential way to achieve calibration data free from blended light is to apply post-processing techniques aimed at source separation. 

Source separation post-processing techniques, such as Principle Components Analysis (PCA) and Independent Components Analysis (ICA), have been successfully applied to a variety of astrophysical cases. Although PCA and ICA both are matrix factorization methods, PCA transforms a reference matrix into linear combinations of vectors, the principle components, that best describe the variance of the data, while ICA transforms a reference matrix into combinations of vectors that best describe the independent features of the data \citep{CICHOCKI}. PCA and ICA work well in a variety of cases that require separating mixed signals, can have notable issues when applied to astrophysical sources, such as images displaying non-physical, negative artifacts due to the matrix decomposition process \citep[KLIP;][] {soummer_2012}.

Another signal processing method, Non-negative Matrix Factorization (NMF), has been explored to decompose matrices when non-negativity is desired to physically interpret data. NMF has been used to reduce artifacts in other astrophysical cases, such as when characterizing circumstellar disks \citep{ren_2018}. A key feature of NMF is its non-negativity, which may be more uniquely suited to eliminate components associated with blended sources or crosstalk between lenslet columns because all signals associated with astrophysical sources are positive.

In this work, we utilize NMF as a way to separate blended spectra apparent in calibration scans obtained from the OH Suppressing InfraRed Imaging Spectrograph (OSIRIS; \cite{larkin_2006}). We apply NMF to calibration methods to test the ability of NMF to mitigate crosstalk between lenslet columns in processed images.

In~\S\ref{sec:NMF_cal}, we describe methods used for calibrating OSIRIS, Non-negative Matrix Factorization (NMF), and our process of applying NMF to calibration scans. Section~\ref{sec:results} presents the impact NMF has on lessening crosstalk between lenslet columns in reduced calibration images and how to optimize applying NMF to calibration data. In~\S\ref{sec:c_and_s}, we conclude by summarizing our results and providing suggestions for improving calibration methods utilized in the existing OSIRIS DRP and the potential NMF application has to other instrument data reduction pipelines. 

\section{Applying Non-negative Matrix Factorization to Calibration Methods}
\label{sec:NMF_cal} 
In this section, we review the various components needed to apply NMF to calibration images. In~\S\ref{sec:cal_and_extract}, we explain how to calibrate and extract spectra from OSIRIS data. Next, in~\S\ref{sec:NMF}, we explain the properties of NMF. Finally, we describe the calibration data collected and how we apply NMF to this data in~\S\ref{sec:experiment}.

\subsection{OSIRIS Calibration and Extraction}
\label{sec:cal_and_extract}
Calibration scans are transformed into a rectification matrix -- a calibration tool specific to OSIRIS that maps the flux associated with the trace on detector to specific pixels on the detector. Each discrete order of the trace is associated with a specific lenslet in a single lenslet column. 

During the calibration process, the lenslet array is flood-illuminated by white light and an exposure is taken at each column position. A physical mask is used to isolate a single lenslet column and the mask’s position can be adjusted along the lenslet array to illuminate each lenslet column individually, as described in~\S\ref{sec:intro}. The light from the illuminated lenslet column creates a trace on the detector that can be used to characterize the flux from each of the lenslets in that column. The flux on the detector is only a function of the lenslet and pixel response since the white light source is uniform. 

We create a rectification matrix from a combination of flood-illuminated images unique to each of the 51 lenslet columns and measure the response of the pixels on the detector for each independent lenslet \citep{lockhart2019}. Rectification matrices are used to reduce raw data by assigning the flux on a two-dimensional detector to a three-dimensional data cube consisting of both spatial and spectral dimensions. The output data cube provides measures of flux for discrete wavelengths and spaxels, or spatial pixels.

Rectification matrices are used in the spectral extraction process to calibrate and deblend spectra. To produce properly deblended spectra, the white light exposures used to create the rectification matrix should not be blended.

Although spectral extraction techniques vary depending on the instrument, the issue of how to most effectively deblend spectra remains common. In comparison to OSIRIS, the data reduction pipelines of other IFSs, such as SAURON, GPI, and CHARIS, do not use the equivalent of a rectification matrix. Instead, they combine measured PSFs of individual lenslets and knowledge of the detector properties to create a global instrument models \citep{bacon_2001, in_2014, brandt_2017}. The reconstruction of lenslet PSFs gives flexibility in modeling PSF aberrations and can also be used to model and remove sources of error, such as crosstalk. Currently, OSIRIS does not have a well-developed PSF model to enable such techniques.
 
\subsection{Non-Negative Matrix Factorization}
\label{sec:NMF}
To separate blended sources caused by light leaking from adjacent lenslet columns, we employ a mathematical factoring method, NMF, to disentangle overlapping signals. Matrix factorization decomposes a primary matrix into various components. In our case, we have a set of vectors, $\vec{x}_j$, that represent the set of possibly blended calibration scans that we would like to transform into un-blended vectors, $\vec{h}_j$.

NMF decomposes a reference matrix, \textbf{X}, consisting of blended calibration scan vectors, $\vec{x}_j$, into both a factor matrix, \textbf{H}, and weight matrix, \textbf{W}. The combination of the factor and weight matrix, \textbf{WH}, contains no negative elements and approximates the original reference matrix, \textbf{X}. 
Given \textbf{W} and \textbf{H}, we minimize the metric $V$ using the following relation:
\begin{equation}
    V(\textbf{W},\textbf{H})=||\textbf{X}-\textbf{WH}||^{2}_{F},
\end{equation}
where divergence is measured using the Frobenius norm. The metric quantifies how similar \textbf{WH} is to the original matrix \textbf{X}. There exist algorithms to solve for the matrix \textbf{WH} which we will not detail here, but are described in \cite{nmf_ref}. 

\subsection{Applying NMF to Calibration Scans}
\label{sec:experiment}

\subsubsection{Data Acquisition}
To create a rectification matrix, one image per lenslet column in needed to  characterize the flux associated with each individual lenslet. For routine calibration sequences, the lenslet column mask is shifted over one lenslet column at time. However, the motor driving the lenslet column mask is capable of taking finer steps and can position the mask between lenslet columns. The flexibility in position of the lenslet mask allows for us to take multiple images per lenslet column. 

Since each calibration image corresponds to a single component of our NMF matrix, we believe increasing the number of components applied to NMF comprised rectification matrices could potentially improve the deblending process.

We obtained a finely sampled data set where the column mask was moved 10 steps across the lenslet array and then an exposure was taken. There are 46--47 motor steps per lenslet position during normal operation, allowing up to $\sim$5 samples to be taken per lenslet position. 
 
\begin{figure*}
\begin{centering}
\includegraphics[scale=1.1]{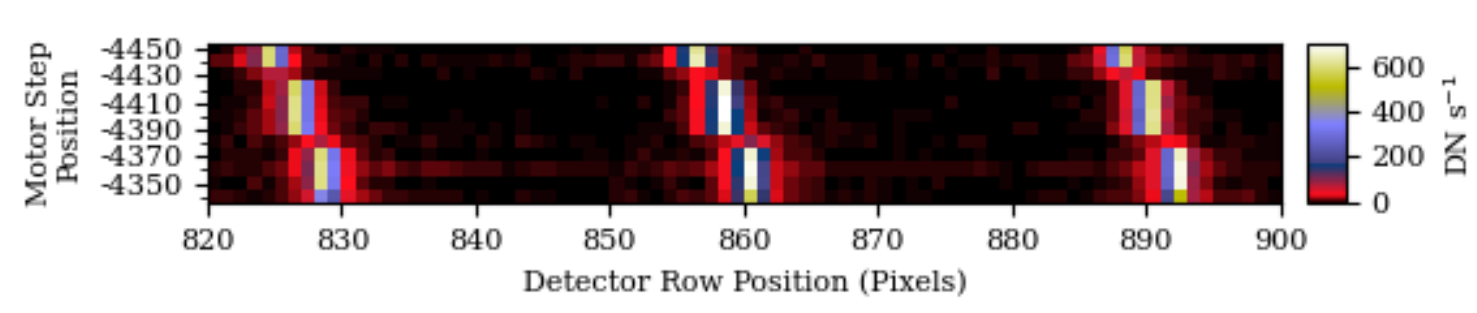}
\caption{The counts on the detector as a function of both the motor step position of the single column lenslet mask and detector row position. The counts associated with each pixel change depending on the position of the single column lenslet mask, showing that the position of the mask impacts what pixel position on the detector the flux is associated with. The motor step position range chosen includes lenslet columns 20, 21, and 22 at the motor step positions of -4447, -4400, and -4353 respectively. These motor step positions are used for calibration scans and leave the lenslet column at that position uncovered. This figure shows a sample range between 820 and 900 pixels across the detector, although position of the lenslet column mask changes the counts associated with individual pixels across the full detector.}
\label{fig:mp_detect}
\end{centering}
\end{figure*}

Figure~\ref{fig:mp_detect} shows how the position of light incident on detector changes depending on the position of the column mask. Looking at a fine motor step position, taken every ten steps, the pixel position of where the light falls on the detector changes based on position of the column mask. If the column mask is not perfectly lined up with a single lenslet column, then the pixel position of the flux associated with a lenslet could be characterized improperly. In order to combat this effect, we process these potentially blended calibration scans using NMF.

\subsubsection{Application}

We utilize an NMF package \citep["sklearn.decomposition.NMF,"][]{scikit} to apply NMF to white-light calibration images. 
First, we subtract darks from the calibration images and flatten them into one-dimensional arrays to create the potentially blended scan vectors, $\vec{x}_j$, that constitute the original reference matrix, \textbf{X}. In our data set, we did not need to correct for elements such as bad pixels or cosmic rays. To account for these in future experiments, we could flag bad pixels and cosmic rays to remove them from the data set, or combine multiple exposures taken at the same motor step position to effectively smooth over cosmic rays at the expense of increasing calibration time.

The number of input vectors, $\vec{x}_j$, changes as a function of the mask sampling density and determines the number of NMF factors to compute. We created NMF factors using a mask sampling density between 1 and 5 scans per lenslet column, or created NMF processed images using 51 to 255 NMF factors. To reproduce the features of an image, NMF finds independent components that, in our case, correspond to light from distinct lenslet columns. The NMF factors are applied to the input vectors, $\vec{x}_j$, to create deblended scan vectors, $\vec{h}_j$. The deblended scan vectors are then combined to create the factor matrix, \textbf{H} and flux is normalized using the weight matrix, \textbf{W}\footnote{A tutorial of how to apply NMF to white-light calibration images can be found at \url{https://github.com/KHorstman/OSIRIS_NMF_tutorial/}}. We produced multiple NMF processed images, \textbf{WH}, by varying numbers of NMF factors. We then created different rectification matrices based on various sampling densities and compared our NMF-derived rectification matrices to an unprocessed rectification matrix. 

\section{Results}
\label{sec:results}
We applied a standard pipeline-processed rectification matrix as well as NMF rectification matrices of different mask sampling densities to arc-lamp exposures of Argon, Neon, Krypton, and Xenon. We compare metrics of error-to-signal, a measure of crosstalk, and signal-to-noise to determine if utilizing NMF in calibration procedures reduces crosstalk between lenslet columns in reduced calibration images.

\subsection{Impact of NMF on Crosstalk between Lenslet Columns}
\label{sec:NMF_cross}

\begin{figure}
\begin{centering}
\includegraphics[scale=0.5]{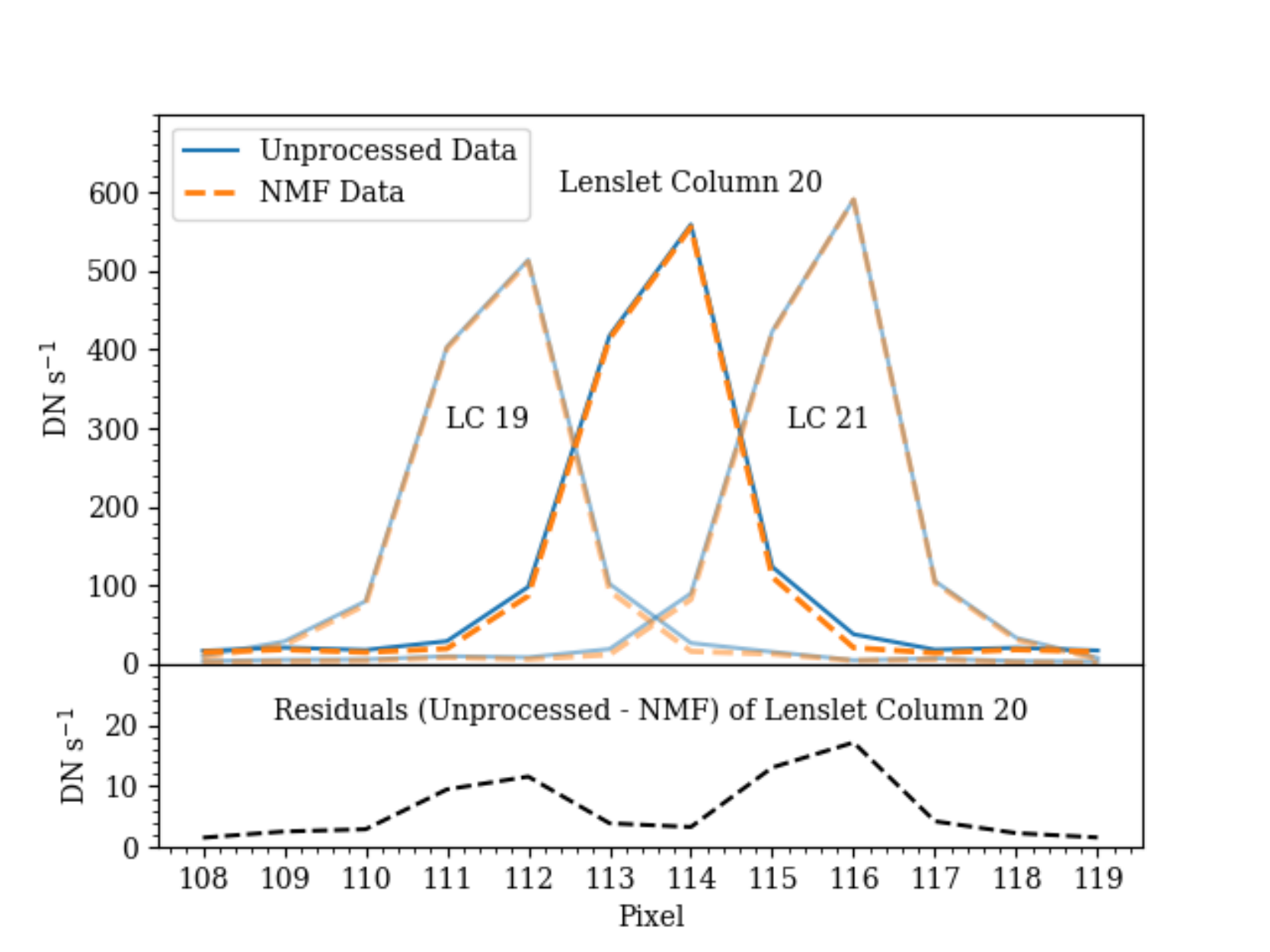}
\caption{\textbf{Top:} Flux variation across one of the traces in a raw image used to construct the rectification matrix. In blue is the flux across the trace of the standard pipeline-processed image and in orange is the flux across the trace after NMF was applied. \textbf{Bottom:} The residuals, or the unprocessed flux minus the NMF flux, of the middle lenslet column, lenslet column 20. NMF reconstruction of lenslet column 20 shows reduced flux in the wings of line profile corresponding to where light from lenslet columns 19 and 21 potentially contribute lenslet column 20's overall flux. Note that all residuals are positive.}
\label{fig:1D}
\end{centering}
\end{figure}

\begin{figure*}
\begin{centering}
\includegraphics[scale=.7]{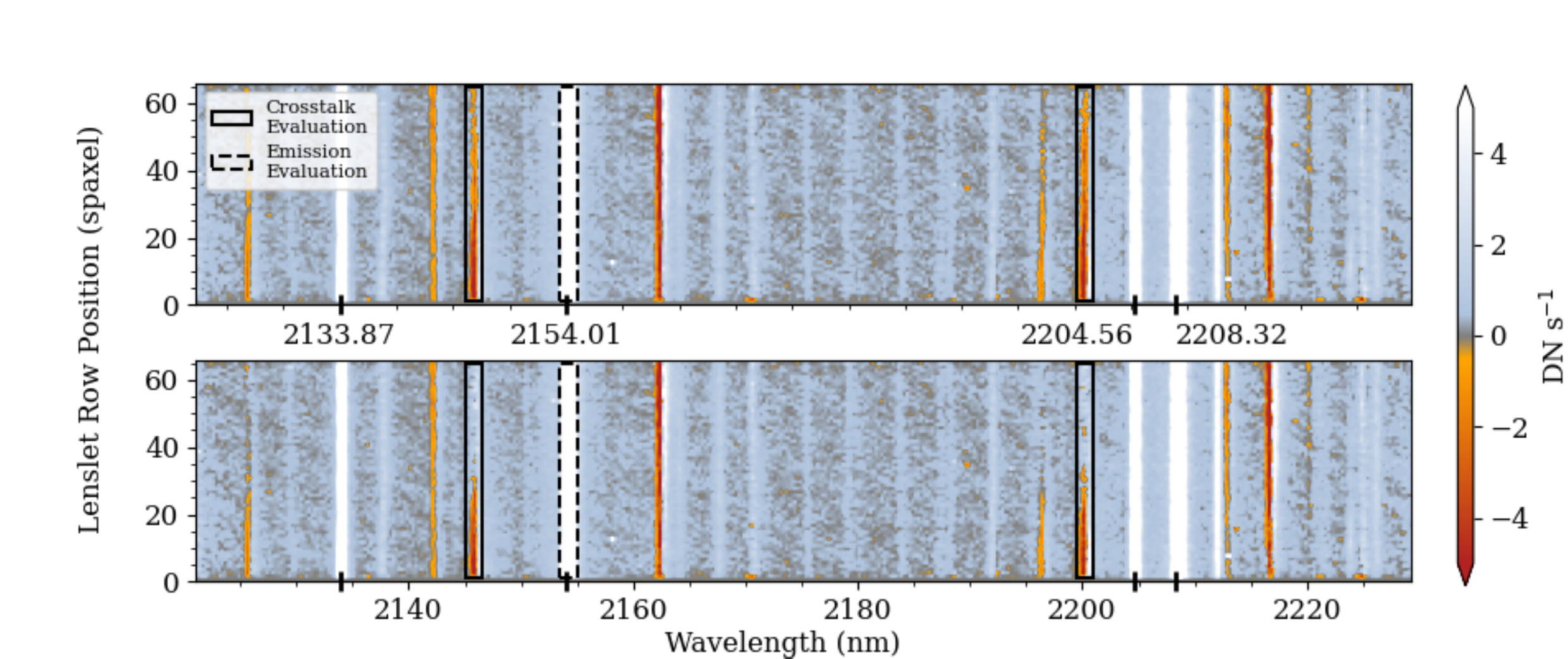}
\caption{Comparison of the reduced image of Argon spectral lines using an unprocessed rectification matrix as well as a rectification matrix created using NMF factors. A two-dimensional slice of the three-dimensional data cube was chosen in such a way to highlight crosstalk. Each image was taken from the same slice, associated with the detector position of lenslet column 20, of the spectral cube. Regions that highlight example areas that correspond to crosstalk between lenslet columns are outlined with solid black lines, while regions that highlight example areas of emission are outlined with dotted black lines. The color map located on the right of the figure indicates key features of the reduced images. The background, shown in gray, is centered on the median absolute deviation of the images. The color map transitions to blue at a value of positive 4$\sigma$ and to orange at a value of negative 4$\sigma$.  \textbf{Top:} Image of Argon spectral lines reduced from an unprocessed rectification matrix. The labels and bold tick marks on the x-axis indicate the wavelengths associated with bright, Argon emission lines. \textbf{Bottom:} Image of Argon spectral lines reduced from a rectification matrix created using NMF factors. Areas of crosstalk outlined in solid black lines appear to be reduced when using the NMF rectification matrix compared to the unprocessed rectification matrix.}
\label{fig:con_NMF_plot}
\end{centering}
\end{figure*}

Figure~\ref{fig:1D} shows how the NMF reconstruction of a lenslet column has reduced flux in the wings corresponding to the locations where adjacent lenslet columns could potentially contribute additional flux. We compare the flux across the trace of an unprocessed scan to NMF factored scan. The flux in the wings of the NMF factored scan is reduced compared to unprocessed data, as shown in the residual plot. This suggests NMF reduces the light leak from other lenslet columns in the wings of the trace where we would expect spectra to overlap. However, a definitive test of the impact NMF has on crosstalk can be measured using arc lamp data. 

Arc lamp data is well-suited to exhibit crosstalk between lenslet columns. Flux mis-assignment in earlier in data reduction process causes crosstalk to appear in reduced arc lamp images as areas of negative flux in conjunction with spectral emission lines. To characterize areas associated with crosstalk, a two-dimensional slice of the three-dimensional data cube was taken at a particular detector position. Slices were taken away from the edges, to ensure the they included no additional artifacts, but still spanned the full wavelength range. The detector position chosen corresponds to the position of a single lenslet column. Figure~\ref{fig:con_NMF_plot} shows an example of the difference in crosstalk between slices of extracted data from an Argon image using both a standard pipeline-processed rectification matrix and an NMF transformed matrix creating using one sample per lenslet column. Qualitatively, we see a reduced spaxel values corresponding to crosstalk in images reduced using the NMF-derived rectification matrix compared to the unprocessed rectification matrix, notably at 2148\,nm and 2220\,nm, marked by solid boxes in Figure~\ref{fig:con_NMF_plot}. We find that the negative flux, or crosstalk, corresponding to negative images of spectral emission lines at different wavelengths is mitigated, while emission regions, denoted by dashed boxes, remains unchanged. We construct quantitative metrics of crosstalk from reduced arc lamp images to measure the reduction in negative flux. 

We aim to mitigate systematic errors, or areas associated with crosstalk, while not impacting the strength of the signal measured or the featureless areas associated with random error, or noise. To analyze if using NMF processed images to create rectification matrices achieves this goal, we define several representative regions within the spectrum.

 For each slice, or lenslet column, we found the signal, random error, and systematic error for an isolated region. The signal is defined as the average value of a spaxel within a region that corresponds to an emission line, denoted in Figure~\ref{fig:con_NMF_plot} as a dashed box. The random error, or noise, is defined as the median absolute deviation of a spaxel in an area with no spectral features. In Figure~\ref{fig:con_NMF_plot}, the area the median absolute deviation was calculated over for each pixel was between 2138 nm and 2141 nm for the full range of 0 to 66 spaxels. The size of the area the median absolute deviation was calculated over varied based on the position and density of emission and crosstalk regions for different arc-lamp images. The systematic error is defined as the average value of a spaxel that corresponds to crosstalk, denoted by the solid boxes in Figure~\ref{fig:con_NMF_plot}. 
 
 After defining several representative regions within the spectrum, we compared metrics of signal-to-noise (signal/random error) and error-to-noise (systematic error/random error) for an image reduced using the unprocessed rectification matrix and an image reduced using an NMF rectification matrix. The final values used to compute the signal-to-noise ratio and the error-to-signal ratio were the median absolute deviation values of signal, noise, and error across all slices of the data cube for each reduced image. 

Next, we analyzed the signal-to-noise and error-to-signal ratios the in reduced images created using both NMF-derived rectification matrices of different mask sampling densities and the unprocessed rectification matrix. Figure~\ref{fig:SN_plot} shows the signal-to-noise ratio and the error-to-signal ratio, respectively. The uncertainty associated with both signal-to-noise and error-to-signal was found empirically by calculating the standard error of the mean for each distribution across all slices of the data cube. The signal-to-noise ratio does not vary significantly when comparing the standard pipeline-processed rectification matrix to NMF-derived rectification matrices, but the error-to-signal is significantly reduced when an NMF-derived rectification matrix is applied. The significance of describing NMF-derived rectification matrices by using mask sampling density as a metric is explained in Section~\ref{opt_NMF}.

\subsection{Optimizing Mask Sampling Density}
\label{opt_NMF}
To produce a rectification matrix, at least one calibration image per lenslet column is needed. 
Taking additional calibration scans scales proportionally with time and is a resource-intensive process, so it is beneficial to optimize the number of scans per lenslet column needed to create a rectification matrix derived from NMF factors. Additionally, there is a potential for the NMF algorithm to more cleanly separate the column responses with more finely sampled mask positions. To balance the number of images per column with the decrease in systematic error, we created several rectification matrices, each with different mask sampling densities as outlined in \S~\ref{sec:experiment}. By comparing both the signal-to-noise and error-to-signal ratios of NMF processed rectification matrices with different sampling densities, we determine the minimum number of white light images needed to produce the most substantial reduction in crosstalk. 

\begin{figure}
\begin{centering}
\includegraphics[scale=0.45]{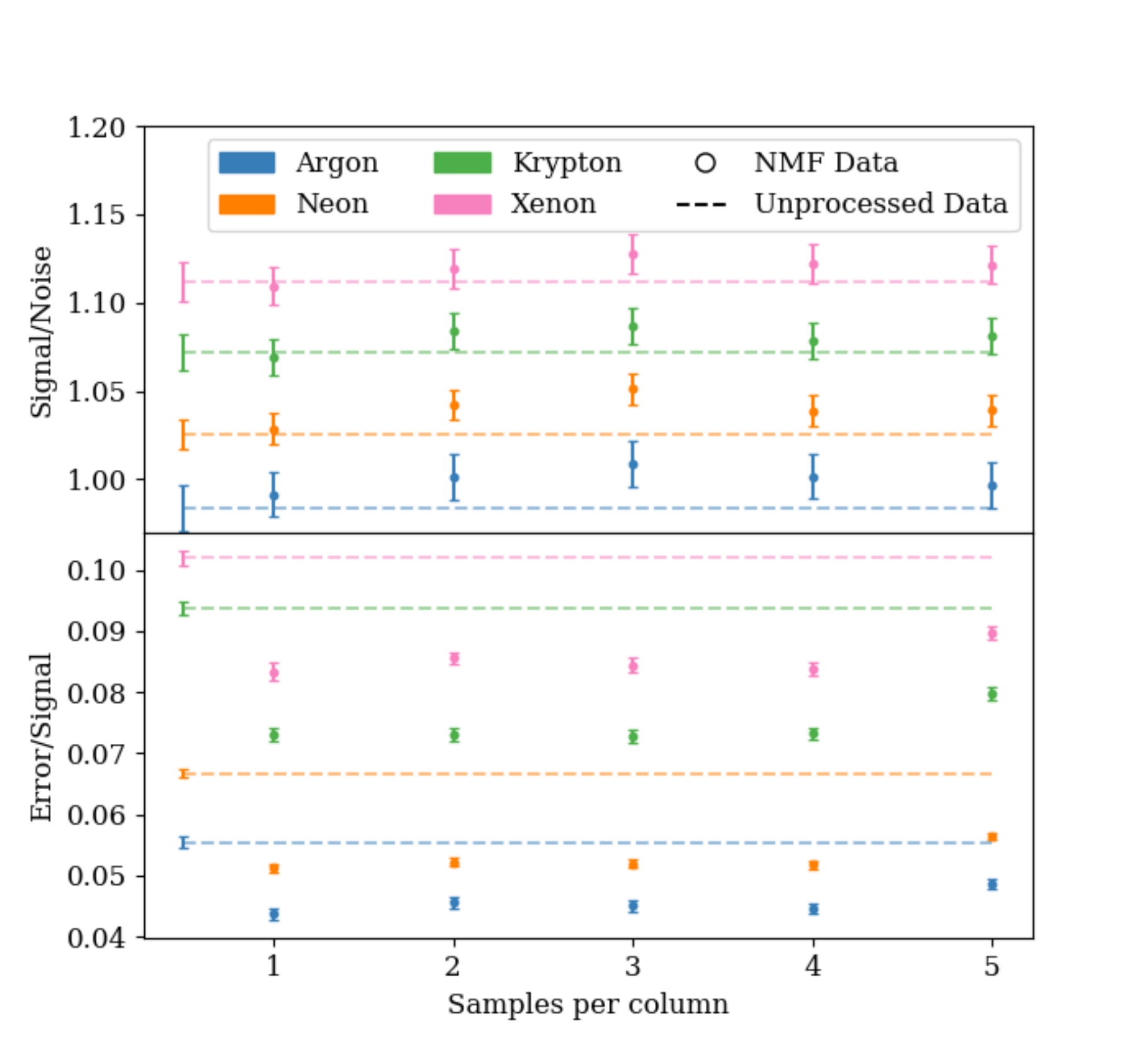}
\caption{Comparison of reduced arc-lamp images using both unprocessed and NMF-derived rectification matrices. \textbf{Top:} Comparison of signal-to-noise for arc-lamp images reduced using an unprocessed rectification matrix and NMF-derived rectification matrices created using different mask sampling densities. The various colors represent different flood-illuminated arc lamps: Argon, Neon, Krypton, and Xenon. The circles represent the signal-to-noise ratio of an image reduced using different NMF-derived rectification matrices, while the dotted line represents the signal-to-noise ratio of an image reduced using a standard pipeline-processed rectification matrix. NMF processing can reduce the signal-to-noise ratio, but overall there is little effect. \textbf{Bottom:} Comparison of error-to-signal, a measure of crosstalk between lenslet columns, for arc-lamp images reduced using a standard pipeline-processed rectification matrix and NMF rectification matrices created using different mask sampling densities. Colors and symbols are the same as in the top panel. The greatest improvement in error-to-signal occurs when 1 sample per lenslet column processed with NMF is used, reducing the error-to-signal by up to 26.7$\pm$0.5$\%$. For all sampling densities and arc-lamp images, NMF processing improves systematic error associated with crosstalk regions.}
\label{fig:SN_plot}
\end{centering}
\end{figure}

Figure~\ref{fig:SN_plot} compares the number of samples per lenslet column used to create NMF processed rectification matrices. In the top panel, the greatest decrease in the signal-to-noise ratio is 0.246$\pm$0.004$\%$ in the Xenon reduced image. The reduction in the signal-to-noise could be caused by the normalization of the NMF factors. For other arc-lamp reduced images, depending on the sampling density, the signal-to-noise ratio increases. The greatest improvement in the signal-to-noise is 2.53$\pm$0.04$\%$. The value of the signal-to-noise ratio using NMF processed rectification matrices deviates little from the value of the signal-to-noise ratio when using the unprocessed rectification matrix. In the bottom panel, the greatest improvement in error-to-signal occurs when 1 sample per lenslet column corresponding to normal calibration procedure, but processed with NMF is used. Using a single sample per lenslet column, crosstalk in the apertures defined above was reduced in calibration images for Argon, Krypton, Neon, and Xenon by 21.1$\pm$0.5$\%$, 26.7$\pm$0.5$\%$, 21.9$\pm$0.4$\%$, and 24.7$\pm$0.7$\%$ respectively. For OSIRIS, it appears that creating rectification matrices from NMF factors using high mask sampling densities does not improve the error-to-signal ratio enough for to justify the use of extra time and additional resources for the acquisition of the extra calibration data. 

Figure~\ref{fig:SN_plot} outlines the benefits of using NMF-derived rectification matrices on data from arc-lamps illuminating the telescope simulator, but to fully understand its potential advantages, we need to apply NMF rectification matrices to on-sky data. However, this is challenging because on-sky and calibration data are obtained differently, complicating our analysis. Even though NMF improves crosstalk between lenslet columns, it does not completely remove it, even in data solely derived from the telescope simulator. When applied to on-sky data, mismatch between the telescope simulator pupil and the on-sky pupil can occur if geometric effects caused by lenslet diffraction and the on-sky pupil dominate the PSF. 
The differences between the PSFs can lead to additional crosstalk. Other sources of potential crosstalk include detector level effects such as inter-pixel capacitance, the brighter-fatter effect, and persistence. For these reasons, NMF rectification matrices may not dramatically improve systematic error when reducing on-sky images when compared to calibration images. 

\section{Conclusions}
\label{sec:c_and_s}
We present our results suggesting Non-negative Matrix Factorization can be used to separate blended spectra taken during calibration scans obtained from OSIRIS. The following motivation and results reflect the viability of using NMF as way to address blended light in calibration images:
\begin{enumerate}
  \item Precise calibration techniques are needed to process raw data from lenslet-based IFS. 
  \item We can treat crosstalk between lenslet columns as a source separation problem, allowing us to apply NMF to calibration scans as a tactic to mitigate unwanted signals.
  \item After analyzing calibration data reduced using NMF-derived rectification matrices, we conclude that NMF factors can mitigate crosstalk between lenslet columns. We find a significant reduction in systematic error, of up to 26.7$\pm$0.5$\%$, corresponding to crosstalk between lenslet columns, but little reduction of the strength of the signal in processed arc-lamp images.
\end{enumerate}

Future work can plan to extend this analysis by reducing on-sky data with NMF rectification matrices. Even though applications to on-sky data may be more complex, we have shown that NMF has utility when applied to calibration scans to potentially improve crosstalk in reduced on-sky images. More generally, our work shows that NMF may be a way to separate blended signals in other IFS data reduction pipelines. 

\section{Acknowledgements}
We thank the referee for a constructive and thorough report. We would also like to thank the OSIRIS working group for their involvement and feedback throughout the duration of this project. 

 The data presented herein were obtained at the W. M. Keck Observatory, which is operated as a scientific partnership among the California Institute of Technology, the University of California and the National Aeronautics and Space Administration. The Observatory was made possible by the generous financial support of the W. M. Keck Foundation. The authors also wish to recognize and acknowledge the very significant cultural role and reverence that the summit of Maunakea has always had within the indigenous Hawaiian community. We are most fortunate to have the opportunity to conduct observations from this mountain. 

\bibliography{ref}

\end{document}